\documentclass[prl,reprint,superscriptaddress,showpacs]{revtex4-1}

\usepackage{amsmath}
\usepackage{amssymb}
\usepackage{amsfonts}
\usepackage{graphicx}% Include figure files
\usepackage{dcolumn}% Align table columns on decimal point
\usepackage{bm}% bold math
\usepackage{lineno}
\nolinenumbers

\begin{document}

\title{Of fishes and birthdays: Efficient estimation of polymer
  configurational entropies}

\author{Ilya Nemenman}
\affiliation{Departments of Physics and Biology, Emory University,
  Atlanta, GA 30322}
\email{ilya.nemenman@emory.edu}

\author{Michael E Wall}
\affiliation{Computer, Computational, and Statistical Sciences}
\email{mewall@lanl.gov}

\author{Charlie E Strauss}
\affiliation{Biosciences Division, Los Alamos National Laboratory, Los Alamos, NM 87545}
\email{cems@lanl.gov}

\date{\today}
\pacs{87.15.A-, 89.70.Cf}

\begin{abstract}
  We present an algorithm to estimate the configurational entropy $S$
  of a polymer. The algorithm uses the statistics of coincidences
  among random samples of configurations and is related to the
  catch-tag-release method for estimation of population sizes, and to
  the classic ``birthday paradox''.  Bias in the entropy estimation is
  decreased by grouping configurations in nearly equiprobable
  partitions based on their energies, and estimating entropies
  separately within each partition.  Whereas most entropy estimation
  algorithms require $N\sim 2^{S}$ samples to achieve small bias, our
  approach typically needs only $N\sim \sqrt{2^{S}}$. Thus the
  algorithm can be applied to estimate protein free energies with
  increased accuracy and decreased computational cost.
\end{abstract}

\maketitle

Computational estimation of protein free-energy differences (e.g.,
between ligand-bound and ligand-free states) is an unsolved problem
with broad applications in molecular biology and medicinal
chemistry. Present approaches to the problem may be divided into two
classes: {\em difference methods}, in which the difference in free
energy between two states is directly estimated, and {\em end-point
  methods}, in which absolute free energies are calculated for the
states being compared.  Difference methods suffer from slow
convergence when there is little overlap between the states, though it
may be possible to overcome this limitation \cite{Ytreberg06}.  In
contrast, end-point methods are independent of the overlap between the
states being compared. They are also trivially more efficient when
pairwise free-energy differences among a large number of states are
required. The main challenge for end-point methods is overcoming
difficulties in estimating configurational entropy, which contributes
substantially to protein free energy
\cite{Frederick-2007,Shirts-2011}. Recent progress has followed
several threads: calibration against reference potentials for which
the free energy can be exactly calculated \cite{Ytreberg06a},
selective sampling about local minima in the energy landscape
\cite{Head97}, hierarchical estimation of chain-elongation transition
probabilities using Monte Carlo simulations \cite{White05}, and
transforming the degrees of freedom to occupy minimally coupled
subspaces \cite{Hensen-2010}. While the calibration method uses only
randomly sampled configurations and their corresponding energies, the
others often require the ability to calculate the energy of an {\em
  arbitrary} configuration.

Entropy estimation has been recognized as a crucial problem in other
disciplines, such as computational neuroscience and cell biology
\cite{Fairhall-2012,Levchenko-2014}.  Properties of entropy estimators
have been studied extensively \cite{Antos01,Paninski03,Zhang12}. Bias,
rather than variance, is the dominant problem. For common estimators,
the bias $\langle \delta S\rangle$ is negative and scales as
\begin{equation}
  \langle \delta S\rangle\equiv\langle S_{\rm est}({\bf n}) -S_{\rm true}({\bf
    p})\rangle \propto- 2^{S_{\rm true}}/N.
\label{bias}
\end{equation}
Here $S_{\rm true}$ is the true entropy of the unknown probability
distribution $\bf p$, $\dim {\bf p}=K$. $S_{\rm est}$ is the entropy
estimated from the measured frequencies $\bf n$, $\sum_{i=1}^Kn_i=N$,
and the averaging $\langle\dots\rangle$ is taken over the random
samples. Both $S_{\rm true}$ and $S_{\rm est}$ are measured in bits.
In particular, the Maximum Likelihood (ML) estimator
\begin{equation}
S_{\rm ML}({\bf n})=-\sum_{i=1}^K \frac{n_i}{N}\log_2 \frac{n_i}{N},
\label{ML}
\end{equation}
which uses the observed frequencies instead of the unknown
probabilities, has bias that scales as in Eq.~(\ref{bias}). This sets
the limit on data requirements for traditional configurational entropy
estimation methods.

When the asymptotic bias follows Eq.~(\ref{bias}), it can be
subtracted from the estimate, making the latter nearly unbiased for
$N\gg 2^{S_{\rm true}}$ \cite{Strong98,Paninski03}. For
$N<2^{S_{\rm true}}$, universally unbiased estimation is impossible
\cite{Antos01,Paninski03}. Then a priori assumptions about the
underlying probability distribution are needed to regularize the
inference. One such Bayesian prior, ${\mathcal P}({\bf p})$, is known
as the NSB (Nemenman-Shafee-Bialek) method \cite{Nemenman02}. It has
been useful in neuroscience, but to our knowledge has not yet been
applied to macromolecular entropy estimation. The approach starts with
noting that seemingly reasonable prior assumptions
${\mathcal P}({\bf p})$ may result in unexpected assumptions
${\mathcal P}(S)$. For example, consider a family of Dirichlet priors
over $\bf p$, indexed by a parameter $\beta$,
\begin{equation}
{\mathcal P} ({\bf p}|\beta) = {1\over Z}\,
\delta\left( 1 - \sum_{i=1}^K p_i\right)
\prod_{i=1}^K p_i^{\beta-1} .
\label{Dirichlet}
\end{equation}
Here the first term normalizes ${\mathcal P}$, and the
$\delta$-function constrains the normalization of the distribution
${\bf p}$ itself. The product of $p_i^{\beta-1}$'s introduces biases
towards peaked ($\beta\to0$) or uniform ($\beta\to\infty$)
distributions $\bf p$. Maximum likelihood inference of ${\bf p}$ with
this prior is equivalent to adding $\beta$ {\em pseudocounts} to every
possible outcome $i$. Importantly, for large $K$, these pseudocounts
bias the resulting Bayesian entropy estimator strongly
\cite{Nemenman02}. The entropy becomes ``known'' before any samples
are measured! One sees this by calculating the a priori entropy
expectation $S_0(\beta)$ and its rms error $\delta S_0(\beta)$ at a
fixed $\beta$, and the latter turns out to be very small
\cite{Nemenman02}. One then uses a new prior over $\bf p$ and $\beta$,
\begin{equation} {\mathcal P}_{\rm NSB} ({\bf p},\beta) = {1\over Z}\,
  \delta\left( 1 - \sum_{i=1}^K p_i\right) \prod_{i=1}^K p_i^{\beta-1}
  \frac{d S_0(\beta)}{d\beta}.
\label{Pflat}
\end{equation}
This is different from Eq.~(\ref{Dirichlet}) by the Jacobian
$d S_0(\beta)/d\beta$, which ensures that, in the limit of a narrow
and monotonic a priori expectation of $S_0(\beta)$, one gets
${\mathcal P}_{\rm NSB}(S_0) = \int d\beta d{\bf p} \,{\mathcal
  P}_{\rm NSB}({\bf p},\beta) \delta (S_0(\beta) - S({\bf p}))\approx
{\rm const}$.
Ref.~\cite{Nemenman02} has argued that this procedure creates a more
uniform ${\mathcal P}(S)$ and reduces the estimation bias.

The NSB estimator is related to the familiar {\em birthday problem}:
in a year with $K$ days, one only needs $N\sim\sqrt{K}$, but not
$N\sim K$, individuals in a room to make it likely that there will be
at least one shared birthday. The same idea is behind the {\em
  catch-tag-release} estimation of wildlife population sizes: in a
pond with $K$ fishes, one will catch a fish that has been previously
caught, tagged, and released after $N \sim\sqrt{K}$ fishes caught. It
follows that one can estimate $K$ by counting how many previously
tagged fishes were caught. If every fish has the same probability to
be caught, then $S=\log_2 K$, and both $S$ and $K$ can be estimated
with $N\sim\sqrt{K}$, compared to the usual methods that require
$N\sim 2^{S_{\rm true}}$, cf.~Eq.~(\ref{bias}). Such estimation of
entropies based on coincidence counting is known as the Ma estimator
\cite{ma-81}. No general estimator can function reliably with fewer
samples: if one never sees a repeat fish, then one only knows the
minimum population size, but nothing about the maximum.

Unfortunately, since logarithms diverge near zero, the
low-probability, poorly sampled tail of $\bf p$ contributes
disproportionally to entropy. Thus coincidence counting cannot
transfer easily to non-uniform probability distributions
\cite{Paninski03}. One needs to use the high-probability events (i.e.,
coincidences) and extrapolate to the tail. Then the prior
${\mathcal P}({\bf p})$ may be seen as enforcing a certain shape of
the tail, and NSB assumes that the tail is not too heavy
\cite{Nemenman02,Nemenman11}.  When the tail structure has been
guessed correctly, the estimate, $S_{\rm NSB}({\bf n})$, converges to
the true entropy in the Ma regime, $N\sim2^{1/2 S_{\rm true}}$
\cite{Nemenman04}. For massive tails, typically there is bias,
$\langle \delta S_{\rm NSB}\rangle<0$. However,
$\left|\langle \delta S_{\rm NSB}\rangle\right|<\left|\langle \delta
  S_{\rm ML}\rangle\right|$.
To verify if a sample size dependent bias is present, one estimates
$S_{\rm NSB}(\alpha N)$, where $0<\alpha\le 1$ is the fraction of data
used. If the estimates at different $\alpha$ agree within the
posterior error bars, the bias can be neglected compared to the
variance \cite{Nemenman08}.

We wondered whether NSB might improve estimation of configurational
entropies of polymer chains. For this, we generated self avoiding
random walks of different lengths on a 3D lattice
\cite{Levitt-1975,Go-1978a,Hinds-1992}. We focused largely on chiral
walks on a cubic lattice, so that the $n$'th bond in the chain is
allowed to take, at most, three of the five possible orientations,
depending on the orientations of the $n-1$'st and the $n-2$'nd
bonds. Such chains weave chiral paths through the lattice,
approximating the c-alpha secondary structure of real proteins
\cite{Karanicolas-2002}. We considered lattices with bounding cubes up
to 4x4x4 (64 total sites) and homopolymeric chains of length
$L \le 50$.  With these constraints, all self avoiding configurations
and their energies can be enumerated on commodity computers. We then
calculated the partition function by direct summation. This makes
further sampling of random configurations trivial and decouples, for
presentation purposes, the entropy estimation problem from the problem
of efficient sampling, which is not the focus of this Letter. To
ensure sufficient generality of our results, we explored chiral chains
of lengths $16\le L\le 50$, as well as short non-chiral chains. The
results were similar for these cases. Thus here we present only chiral
polymers with $L=32$.

In the spirit of Ref.~\cite{Gront-2000}, we evaluated the energy of
lattice conformations using energy functions that include local and
long-range contributions: (1) backbone secondary structure (SS)
propensity, measured by a preference among the 3 chiral local
configurations \cite{Pokarowski-2005}; (2) an approximation of solvent
exposure per residue \cite{Kabsch-1983}, measured by the number of
vacant sites surrounding each occupied lattice site in a fold; and (3)
pair contact energy via a G${\rm \bar{o}}$-like model for preferred
contacts \cite{Go-1978a}. The G$\bar{\rm o}$ model and SS propensities
were derived by simulating arbitrary chains and then arbitrarily
selecting a chain representative of a good protein fold (high contact
order, low solvent exposure, and low radius of gyration)
\cite{Go-1978a,Lau-1989,Dill-1990,Plaxco-1998,Rohl-2004} as a
reference model. We then assigned energies to the other chains in
proportion to their distance from the reference.  We quantified
contact order as the average distance in sequence of two residues
contacting in a fold. Contact order is a key statistic predicting
folding rate, with high values indicating that the fold's nucleation
requires distal regions to come into contact, making it dependent on
meshing of long range side-chain forces in addition to local backbone
propensities \cite{Plaxco-1998}. In the G${\rm \bar{o}}$ model, chains
with the same secondary structure motifs (short range potential) or
the same pairwise long range contacts (not necessarily with the same
neighbors) as the reference model have the lowest energy.  Since the
relative importance of the effects represented by the different energy
functions is unknown a priori, we explored each energy function
independently, arguing that if NSB works for each function, it will
work for their combinations.

The choice of the temperature $T$ for the analysis is very
important. Indeed, for $T\to0$, the configurational distribution is
dominated by a few highly probable configurations. The entropy is low,
and hence it is easy to estimate, cf.~Eq.~(\ref{bias}). For
$T\to\infty$, all configurations are equiprobable, and the Ma
estimator is unbiased when $N\sim \sqrt{K}$, where $K$ is now the
total number of configurations. Intermediate temperatures are the most
interesting. Here entropy is too high for simple methods to work,
while the Ma estimator cannot be used because configurations are not
equiprobable. For our chiral polymers, we estimate numerically
$K\approx 2.74^L$, where 2.74 replaces 3 due to self-avoidance and
finite volume. Thus, for $T\to\infty$,
$S_{\rm true}\sim\log_2 2.74^L$, which is about 46 bits for $L=32$.
We are most interested in entropies substantially smaller than this,
but much larger than 1 bit.

\begin{figure}[t]
  \centerline{\includegraphics[width=2.5in]{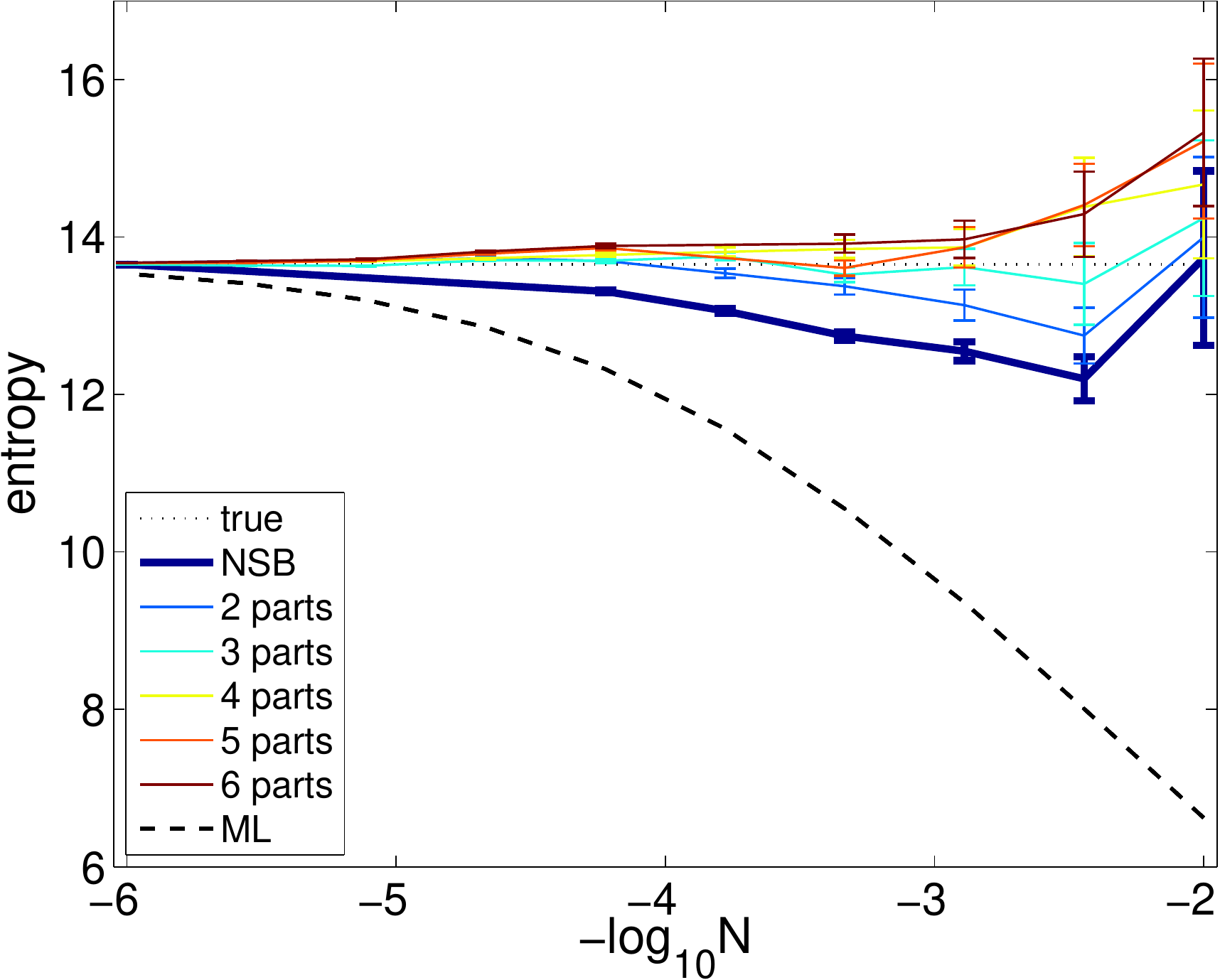}}
  \caption{\label{fig:bias} Configurational entropy estimation for the
    first energy function and $T=1$ a.\ u. NSB (thick line) and the
    grouping estimator with $M=2\dots6$ partitions are shown in
    comparison to $S_{\rm true}$ and the ML estimator. We sampled up
    to $N=10^6\approx2^{20}$ from the Boltzmann distribution. Error
    bars correspond to one posterior standard deviation. For NSB and
    the grouping estimator, we bounded the number of configurations in
    each partition and in total as $\approx 2.74^L$.  The bias of ML
    can be inferred and subtracted out when
    $\log_2N\gtrsim S_{\rm true}\approx14$, so that $N\gtrsim 10^4$,
    which corresponds to the bend in the ML curve
    \cite{Strong98}. However, both ML and NSB are biased for smaller
    $N$. In contrast, the grouping estimators is unbiased for $M=3$ in
    the Ma regime of $\log_2N\gtrsim S_{\rm true}/2\approx7$, or
    $N\gtrsim 10^2$.}
\end{figure}

Typical results of applying NSB to samples from the protein
configurational distribution for the first energy function are
illustrated in Fig.~\ref{fig:bias} at an intermediate temperature
$T=1$ a.\ u., when $S_{\rm true}\approx 13.65$ bits. By the time
$\log_2 N\sim S_{\rm true}/2\approx 7$, many coincidences have
occurred. The estimator is reporting small posterior variances, but it
is biased, though always less than ML. NSB remains biased even when
$N\sim 2^{S_{\rm true}}$. The bias finally disappears only when even
the naive ML estimator is nearly unbiased, $N\gg 2^{S_{\rm true}}$,
that is, many samples per typical configuration. Similar failures are
observed for different sequence lengths and the other two energy
functions. The bias likely stems from the assumptions of NSB being
incompatible with the data.

\begin{figure}[t]
  \centerline{\includegraphics[width=2.5in]{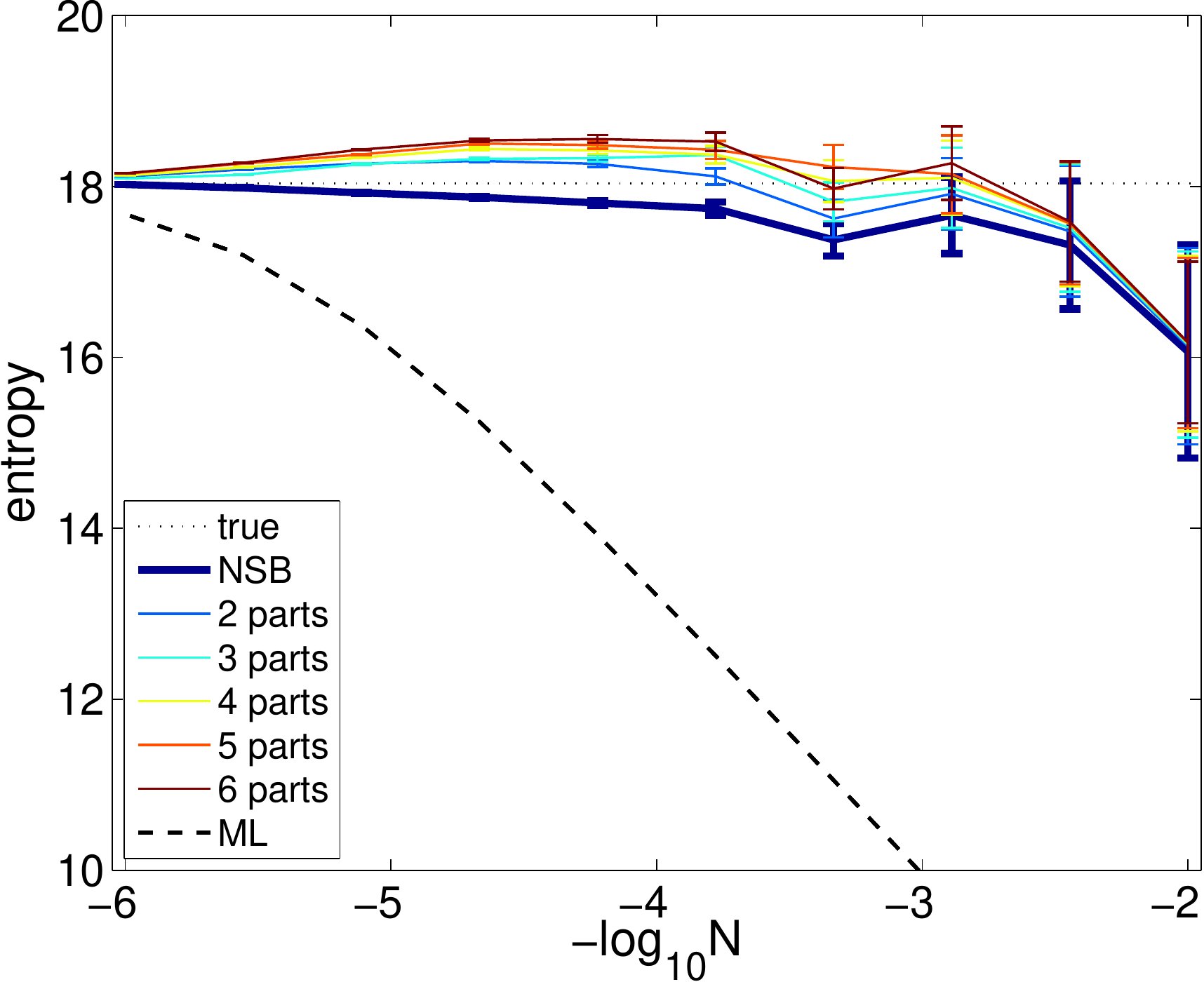}}
  \caption{\label{fig:highT}Configurational entropy estimation for the
    first energy function and $T=3$ a.\ u. Same conventions are used
    as in Fig.~\ref{fig:bias}. Note that the bias is much less of a
    problem for this higher entropy case for NSB. As before, the
    grouping estimator can be made unbiased in the Ma regime,
    $\log_2N\gtrsim S_{\rm true}/2\approx9$, or $N\gtrsim10^3$.}
\end{figure}

As the temperature increases above $T=3$ a.\ u.\, and the entropy
grows beyond $S_{\rm true}\approx 18$ bit, the bias of $S_{\rm NSB}$
becomes small at $\log_2 N > S_{\rm true}/2\approx 9$, see
Fig.~\ref{fig:highT}. While the bias is nonzero, it is comparable to
the standard deviation, making the estimator useable. Thus long tails
disappear from the distribution of configurations, and the estimator
works at $<40$\% of the maximum possible entropy (46 bits) for this
polymer!  Since no general entropy estimator can work until
$\log_2 N \gtrsim S_{\rm true}/2$, in this regime, NSB performs nearly
optimally.

We can capitalize on the accurate performance of NSB for near-uniform
distributions at high $T$. In the fish counting problem, basses,
carps, and catfishes may have different probabilities of being caught,
while the probabilities may be closer to uniform within the
species. Thus counting each species {\em separately} will improve
population estimates, but at the cost of needing a larger $N$ to
ensure that coincidences (catching a tagged fish) happen for each
species. Similarly, suppose the space of possible protein
configurations is split into partitions $\nu_\mu$,
$\mu=1,\dots,M$. Then by the grouping axiom for entropy,
\cite{Shannon62}
\begin{equation}
S({\bf p})=\sum_{\mu=1}^M \pi_{\mu} S(\nu_\mu) + S(\boldsymbol{\pi}),
\label{grouping}
\end{equation}
where $S(\nu_\mu) $ stands for the entropy of the partition $\nu_\mu$,
$\pi_\mu=\sum_{i\in \nu_\mu} p_i$ is the probability of a particular
partition, and $S(\boldsymbol{\pi})$ is the entropy of the partition
choice. While the overall ${\mathbf p}$ may be incompatible with NSB,
the estimator may perform better on each of the partitions separately,
resulting in the new {\em grouping} estimator \footnote{The grouping
  property, Eq.~(\ref{grouping}), was first used to estimate entropy
  in neuroscience context \cite{Nemenman08,Tang-2014}. However, the
  optimal choice of the number of partitions was not addressed
  there.}:
\begin{align}
  S_{\rm gr}({\bf n},M)=&\sum_{\mu=1}^M \phi_{\mu} S_{\rm NSB}(\nu_\mu) +
  S_{\rm NSB}(\boldsymbol{\phi}),\label{grouping_empirical}\\
  \delta^2S_{\rm gr}({\bf n},M)=&\sum_{\mu=1}^M
  \left[\delta^2\phi_{\mu} S_{\rm NSB}(\nu_\mu)+\phi_{\mu} \delta^2
    S_{\rm NSB}(\nu_\mu)\right]\nonumber \\
  &+ \delta^2 S_{\rm NSB}(\boldsymbol{\phi}) \nonumber\\\approx&\sum_{\mu=1}^M
  \phi_{\mu} \delta^2 S_{\rm NSB}(\nu_\mu)+ \delta^2 S_{\rm
    NSB}(\boldsymbol{\phi}).\label{gr_posterior}
\end{align}
Here $\phi_\mu=\sum_{i\in \nu_\mu} n_i$ are the empirical frequencies
of each partition, and $\delta^2$ is the posterior variance. For
$S_{\rm gr}(M)$ to be unbiased, the partitions should be chosen such
that either (i) distributions of configurations within each partition
are more uniform (allowing Ma's arguments to work), or (ii) structure
of tails within each partition is compatible with NSB.  If a non-NSB
estimator is used in the r.\ h.\ s.\ of
Eq.~(\ref{grouping_empirical}), partitions should be chosen instead to
make that estimator unbiased.

Since each polymer configuration has an energy value that is known,
and configurations with similar energies are nearly equiprobable, a
natural partitioning exists in this context.  We expect reduction in
bias if one assigns a configuration with energy $E$ to the partition
$\mu$, for which
$E_{\min}+ (E_{\max}-E_{\min})(\mu-1)/M \le E < E_{\min}+
(E_{\max}-E_{\min})\mu/M$,
where $E_{\min}$ and $E_{\max}$ are the minimum and the maximum energy
in a given sample.

However, such partitioning comes at a cost. First, each of the terms
in Eq.~(\ref{grouping}) has statistical errors. The errors add in
quadratures, so that the estimator variance, $\langle \delta^2 S_{\rm
  gr}\rangle$, typically grows with $M$. Second, when $M \to K$,
$S(\pi)$ approaches $S_{\rm true}$ and becomes equally hard to
estimate. Third, for $M>1$, one needs coincidences in {\em each}
partition. This requires more data, and would lead to $\langle \delta
S_{\rm gr}\rangle>0$ if some of the partitions have no
coincidences. Finally, one doesn't know the maximum possible number of
configurations $K_\nu$ in each partition, and has to take
$K_\nu=K$. Larger $K$ results in a larger $S_{\rm NSB}$, though the
dependence is weak \cite{Nemenman04}. Thus $\langle \delta S_{\rm
  gr}\rangle >0$ for $M\gg 1$. Combined, these concerns indicate that
success of the grouping estimator in polymer problems is uncertain.

We tested the performance of the grouping estimator for different $L$,
$T$, and energy functions. The results were consistent with the
expectations and similar for all cases. As seen in
Fig.~\ref{fig:bias}, increasing the number of partitions first
decreases the bias. $\langle\delta S_{\rm gr}\rangle$ is insignificant
for $M\sim 2\dots 4$. For $M$ so small, each partition is sampled
well, and $S_{\rm gr}$ works in the Ma regime. However, as $M$ grows,
the bias changes sign and increases again.  Similar results hold for
higher temperatures, Fig.~\ref{fig:highT}. Here the bias is small for
all $M$, and it is dramatically smaller than the ML bias.

These results suggest a straight-forward algorithm for estimation of
polymer configurational entropies. For a given sample of
configurations and their energies, one computes
$S_{\rm gr}(\alpha N, M)$ using Eq.~(\ref{gr_posterior}) and
$\delta^2S_{\rm gr}(\alpha N, M)$ for $M=1$ using
Eq.~(\ref{gr_posterior}), while varying the fraction of the data used
for the estimation, $0<\alpha\le 1$. One looks for the sample size
dependent bias by verifying if $S_{\rm gr}(\alpha N, 1)$ drifts by
more than the standard deviation as $\alpha$ increases. If the bias is
positive, the algorithm cannot be applied (this has never happened in
our tests). If the bias is insignificant, then
$S_{\rm gr}(N, 1)\pm \left(\delta^2 S_{\rm gr}(N, 1)\right)^{1/2} $ is
the entropy estimate. If the bias is negative, then one increments
$M\to M+1$, and repeats the estimation for various $\alpha$. One
increments $M$ until it reaches $M^*$, such that
$ \delta S_{\rm gr}(N, M^*) >0$. The best estimate, the bias, and the
variance are then the means of the corresponding quantities for
$S_{\rm gr}(N, M^*-1)$ and $S_{\rm gr}(N, M^*)$.  Crucially, unless
$M^*\gg1$, coincidences are present in all partitions. Thus the
proposed estimator will work in the Ma regime,
$\log_2 N\sim S_{\rm true}/2$, providing a {\em square root} data
requirement reduction compared to simpler approaches.  Since
coincidences are required for any estimator to work, it is unlikely
that other general purpose estimators will substantially outperform
the NSB-based grouping algorithm.

For off-lattice polymers, the entropy can be computed by enumerating
local minima in the energy landscape and additionally estimating the
entropy within each such basin of attraction \cite{Ming-2008}.  For
the latter, there are good methods for entropy estimation based on
kernel smoothing or nearest neighbor techniques
\cite{Beirlant97,Kraskov04}; we expect that these also will be
improved by grouping. Alternatively, the entropy in a local basin may
be estimated analytically using the normal modes approximation
\cite{Ming-2008}. We expect the present version of the NSB algorithm
with grouping to be especially useful for the former, that is for
calculating contributions to entropy from many similar local minima,
which are observed for rugged energy landscapes that are
characteristic of real proteins \cite{Frauenfelder-1991}.

In summary, in this Letter, we have verified that the grouping
generalization of the NSB algorithm can be used to produce reliable
estimates of configurational entropies of polymer chains in the
severely undersampled regime $N\sim\sqrt{2^{S_{\rm true}}}$, using
only random samples of configurations and their corresponding
energies.  The estimator is available from
\url{http://nsb-entropy.sourceforge.net} as a C++ and Matlab/Octave
code. The estimation is rapid on modern computers.  For lattice
polymers of length $\sim30$, it requires only $\sim1000$ configuration
samples. Thus the square-root scaling suggests that the method will be
able to work with sequences of previously unaccessible lengths.

\begin{acknowledgments}
  This work was partially supported by the US Department of Energy
  under the contract No.\ DE-AC52-06NA25396. IN was partially
  supported by the James S.\ McDonnell Foundation Award No.\
  220020321.
\end{acknowledgments}

%\bibliography{Entropy}

\end{document}